\documentclass[3p,times,procedia]{elsarticle}
\usepackage{nupha_ecrc}

\volume{00}

\firstpage{1}

\journalname{Nuclear Physics A}

\runauth{S. Campbell}

\jid{nupha}

\jnltitlelogo{Nuclear Physics A}

\usepackage{amssymb}

\usepackage[figuresright]{rotating}

\begin{document}

\begin{frontmatter}



\dochead{XXVIth International Conference on Ultrarelativistic Nucleus-Nucleus Collisions\\ (Quark Matter 2017)} 

\title{Photon and dilepton observables: An experimental overview}


\author{S. Campbell}

\address{Columbia University Pupin Hall, 5285  538 W 120th St New York, NY 10027}

\begin{abstract}
Dilepton and direct photon measurements in heavy ion collisions provide access to the early stages of the collision.  Dilepton continuum measurements at SPS and RHIC show an excess consistent with broadening of the rho meson spectral function as a result of hadronic interactions.  Direct photon measurements at RHIC and the LHC show an excess of low $p_{T}$ photons over the expected yield.  This excess has a large azimuthal anisotropy with respect to the collision plane.  Understanding the direct photon excess and its anisotropy has become an outstanding question in the field of heavy ion physics.  This proceeding consists of a summary of new dilepton and direct photon experimental results shown at the 2017 International Conference on Ultrarelativistic Nucleus-Nucleus Collisions and presents them in the context of our current understanding of these electromagentic probes.

\end{abstract}


\end{frontmatter}


\section{Introduction}
\label{Sec:Intro}

In an ultrarelativistic heavy ion collision, the initial hard scattering is followed by a pre-equilibrium phase before the Quark Gluon Plasma (QGP) forms and thermalizes.  As the plasma cools and expands, it becomes a hot gas of interacting hadrons until the hadrons cease to interact amongst themselves and free-stream.  Direct photons and dileptons are generated throughout the collision evolution and do not undergo any secondary interactions once emitted.  As a result, they provide clear probes of early stages of the collision.  In particular, prompt and pre-equilibrium photons are created in the initial hard interaction and pre-equilibrium stages respectively, and thermal photons are produced in both the QGP and hadron gas stages.  Similarly, the dilepton spectrum can reveal the in-medium electromagnetic spectral function and possible modifications due to chiral symmetry restoration.~\cite{Rapp2}  In the hadron gas stage, dileptons are generated from virtual rho production via hadronic interactions such as pion-pion annihilation.  Dileptons from rho meson decay can show modifications to the rho meson spectral function from baryon interactions.

New experimental measurements of dileptons and direct photons shown at the 2017 International Conference on Ultrarelativistic Nucleus-Nucleus Collisions (Quark Matter 2017) are presented in these proceedings in two sections.  First the dilepton continuum measurements are shown in Section~\ref{Sec:Dilept}, then the direct photon measurements are shown in Section~\ref{Sec:DirPhot}.  Finally, conclusions and acknowledgements are presented in Sections~\ref{Sec:Concl} and~\ref{Sec:Ackno} respectively.

\section{Dilepton continuum measurements}
\label{Sec:Dilept}
Previous measurements of the dilepton continuum mass spectrum at SPS and RHIC show an excess in the vicinity of the rho meson mass.  This was first measured at the SPS with the dimuon mass spectrum by the CERES experiment in 200 AGeV $S$+$Au$ collisions~\cite{CERES} and then more precisely measured by the NA60 experiment in $158$ AGeV $In$+$In$ collisions.~\cite{NA60}  For masses above $1$ GeV/$c^2$, NA60 measured a thermal-like mass spectrum that can be described by a Boltzman distribution extracting an inverse slope, also called an effective temperature $T_{Eff}$.~\cite{NA60therm}  At RHIC, the PHENIX~\cite{PHENIXee} and STAR~\cite{STARee} experiments have measured the dielectron mass spectrum in $\sqrt{s_{NN}} = 200$ GeV $Au$+$Au$ collisions.  Again, the spectra showed modifications from the expected yields in the vicinity of the rho meson mass.  To identify these modifications the dielectron mass spectra are compared to cocktails of the known hadronic sources, either by subtracting the cocktail as in the case of NA60 or in the ratio of the data over the cocktail as is done at PHENIX and STAR.  The excesses in the dilepton spectra relative to these cocktails are well-described by the same theory for SPS and RHIC energies.~\cite{Rapp}  This model finds the excess is due to the broadening of the rho meson's spectral function from interactions in the hadron gas phase. New and future dilepton measurements from the HADES, ALICE, STAR and PHENIX experiments are described in the remainder of this section.

The HADES experiment presented preliminary results of the dielectron mass spectrum in $\sqrt{s_{NN}}=2.4$~GeV $Au$+$Au$ collisions at GSI.~\cite{HADESdie}  The baryon-rich environment created at these lower collision energies provides a new avenue to understand rho meson production and modification, in particular rho modifications due to baryon interactions.  The measured dielectron spectrum is compared to a reference distribution from $p$+$p$ and $p$+$n$ collisions.  The dielectron excess is isolated when the reference spectrum is subtracted with the eta contribution from the continuum measurement.  Both the dielectron mass spectrum with and without the eta and $\frac{1}{2}(np + pp)$ reference spectrum subtracted were shown.  
The resulting excess spectrum is well described by coarse-grained transport models with rho-baryon coupling.~\cite{CGgsi}~\cite{CGfra}  The dielectron excess can also be described by a Boltzman distribution with an inverse slope of $70.67 \pm 4.44$~MeV/$k_{B}$.  This HADES result reveals the behavior of rho meson production at high net baryon density, extending our understanding of dilepton production to new areas of the QCD phase diagram.

\begin{figure}
\centering
\begin{minipage}[t]{7.2cm}
  \centering
  \includegraphics*[width=7cm]{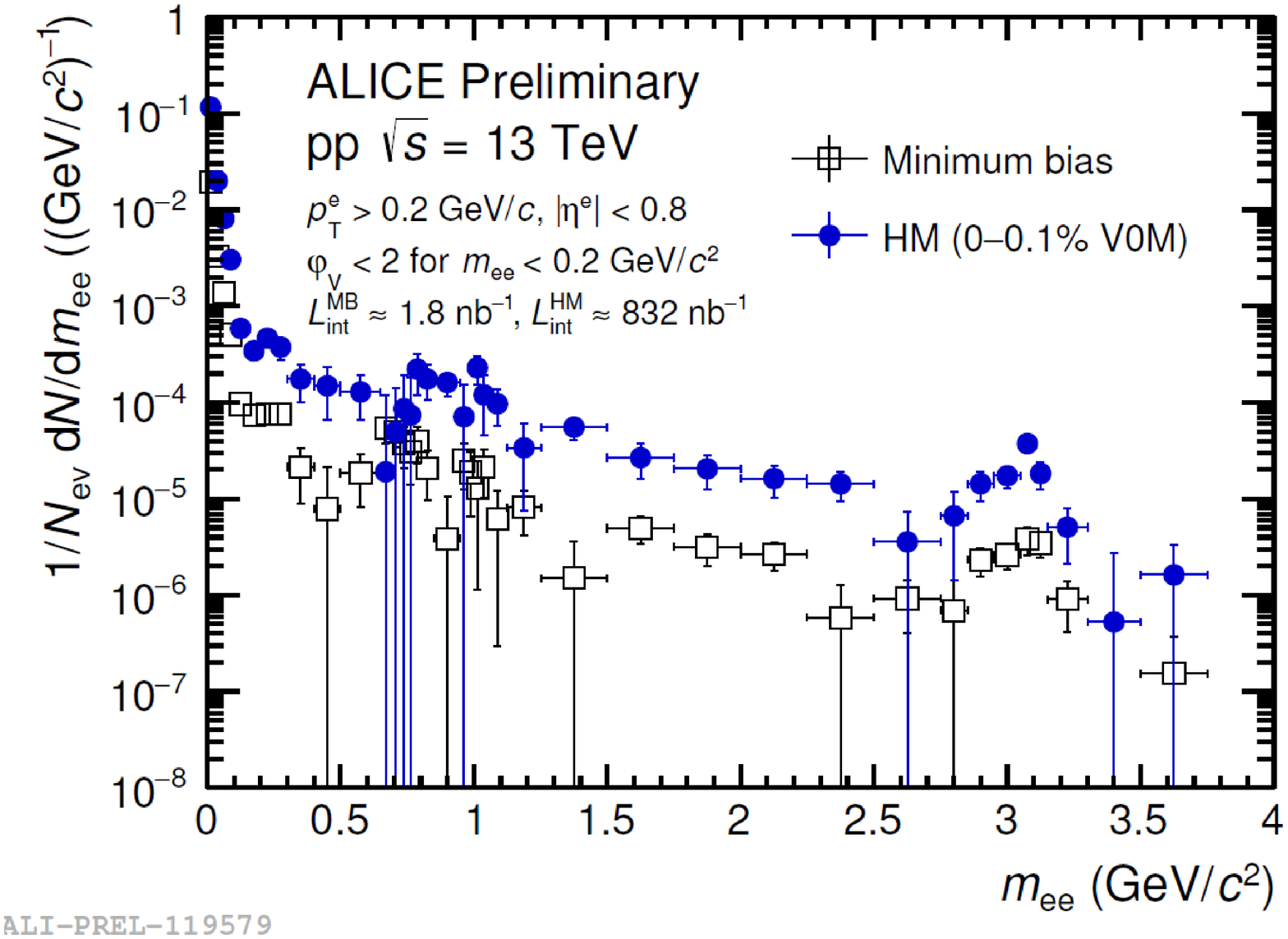}
  \label{Fig:ALICEee_highmult}
\end{minipage}
\hfill
\begin{minipage}[t]{7.2cm}
  \centering
  \includegraphics*[width=7.cm]{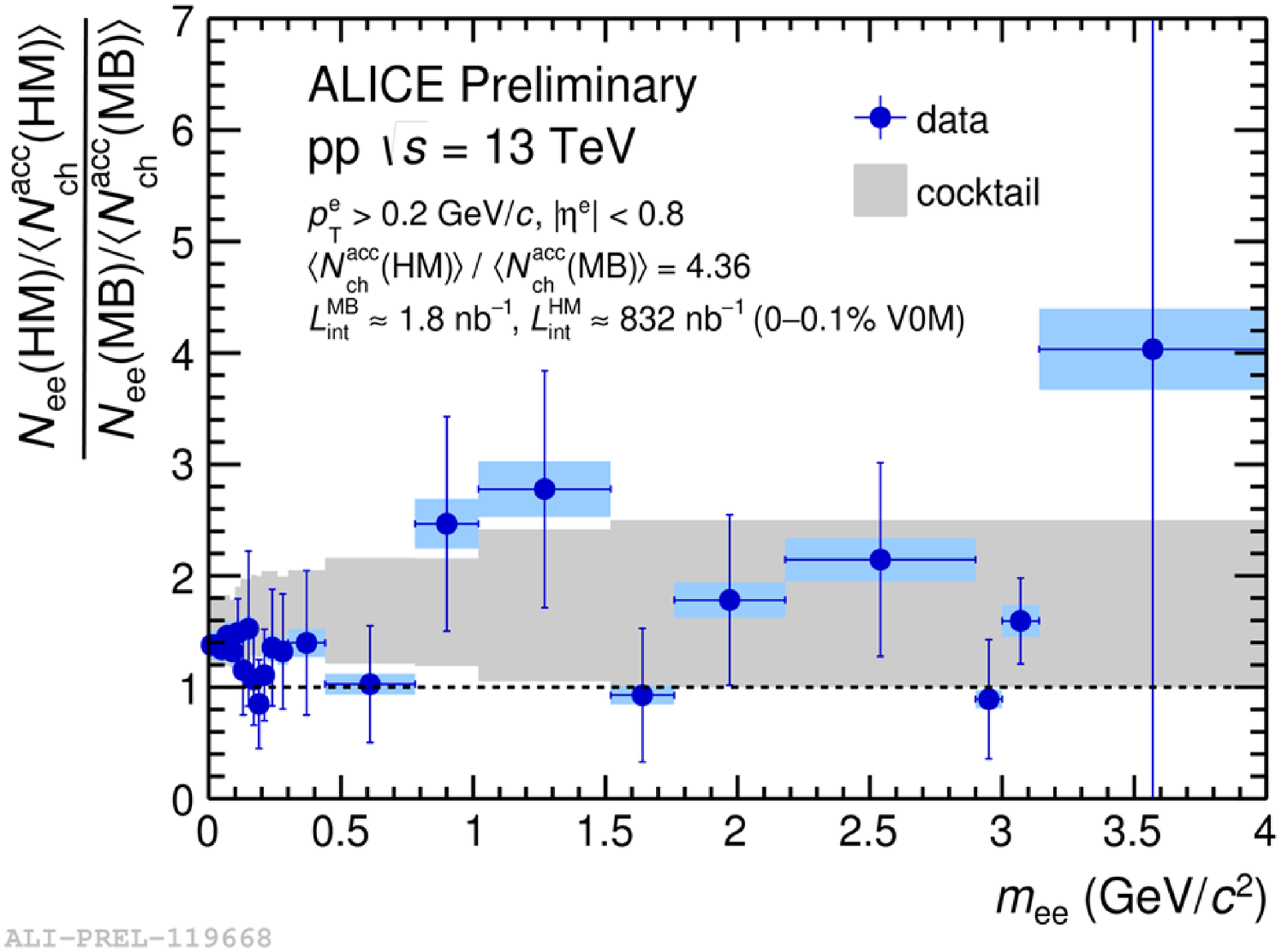}
  \label{Fig:ALICEeeratio_highmult}
\end{minipage}
\caption{On the left are preliminary results from the ALICE collaboration of the raw dielectron mass spectra in $\sqrt{s_{NN}}=13$~TeV $p$+$p$ collisions.  These spectra are for high multiplicity triggered (blue closed) and minimum bias (black open) events.  On the right is the ratio.}
\label{Fig:ee_highmult}
\end{figure}

The ALICE collaboration showed a preliminary measurement of the dielectron continuum mass spectrum in high multiplicity $\sqrt{s_{NN}}=13$~TeV $p$+$p$ collisions at the LHC.~\cite{ALICEdie}  The high-multiplicity events are identified by the high-multiplicity trigger and have a charged particle yield increased by a factor of 4.35 relative to the minimum bias $13$~TeV $p$+$p$ yield.  Similarly, the dielectron spectrum in high-multiplicity events has an increased per event yield compared to the minimum bias dielectron spectrum.  The ratio of the multiplicity-scaled raw dielectron spectrum in these high multiplicity events over the multiplicity-scaled minimum bias $p$+$p$ spectrum is greater than one.  Figure~\ref{Fig:ee_highmult} shows the high multiplicty and minimum bias dielectron mass spectra on the left and the multiplicity-scaled ratio of the high multiplicity spectrum over the minimum bias spectrum on the right.  The greater than unity behavior is also seen in the ratio of the hadronic cocktail contributions and may be a result of the increase in the mean $p_{T}$ in high multiplicity $p$+$p$ events.  The high multiplicty $p$+$p$ analysis is a promising new measurement that will benefit from reduced statistical and systematic error bars when the full dataset, with a factor of five increase in the sample size, is analyzed.

The STAR experiment presented new preliminary results investigating the dielectron continuum in $\sqrt{s_{NN}}=200$~GeV $Au$+$Au$ collisions at low $p_{T}$ of the dielectron pair, $p_{T} < 0.15$~GeV/$c$.~\cite{STARdie}  They found a large excess at all masses that was particularly clear in the peripheral centrality bin.  Coherent photo-production was proposed as a potential source of this excess.  The mass distribution of the excess does not contain a rho-like mass peak, suggesting a larger contribution from photon-nuclear interactions instead of photon-photon interactions if the source is indeed coherent photo-production.  Such an excess is not seen in the $p_{T} < 0.5$~GeV/$c$ PHENIX spectrum in minimum bias $d$+$Au$ collisions.~\cite{PHENIXdAudie}  However this could be a result of the reduced rates of photo-production expected in $d$+$Au$ collisions relative to $Au$+$Au$ collisions, a difference of a factor of roughly $Z$ in $d$+$Au$ versus a factor of $Z^2$ in $Au$+$Au$.  A comparison to a similarly low $p_{T}$ peripheral heavy ion dilepton measurement from PHENIX or the LHC experiments is needed.

\begin{figure}
\centering
\begin{minipage}[t]{7.2cm}
  \centering
  \includegraphics*[width=6.4cm]{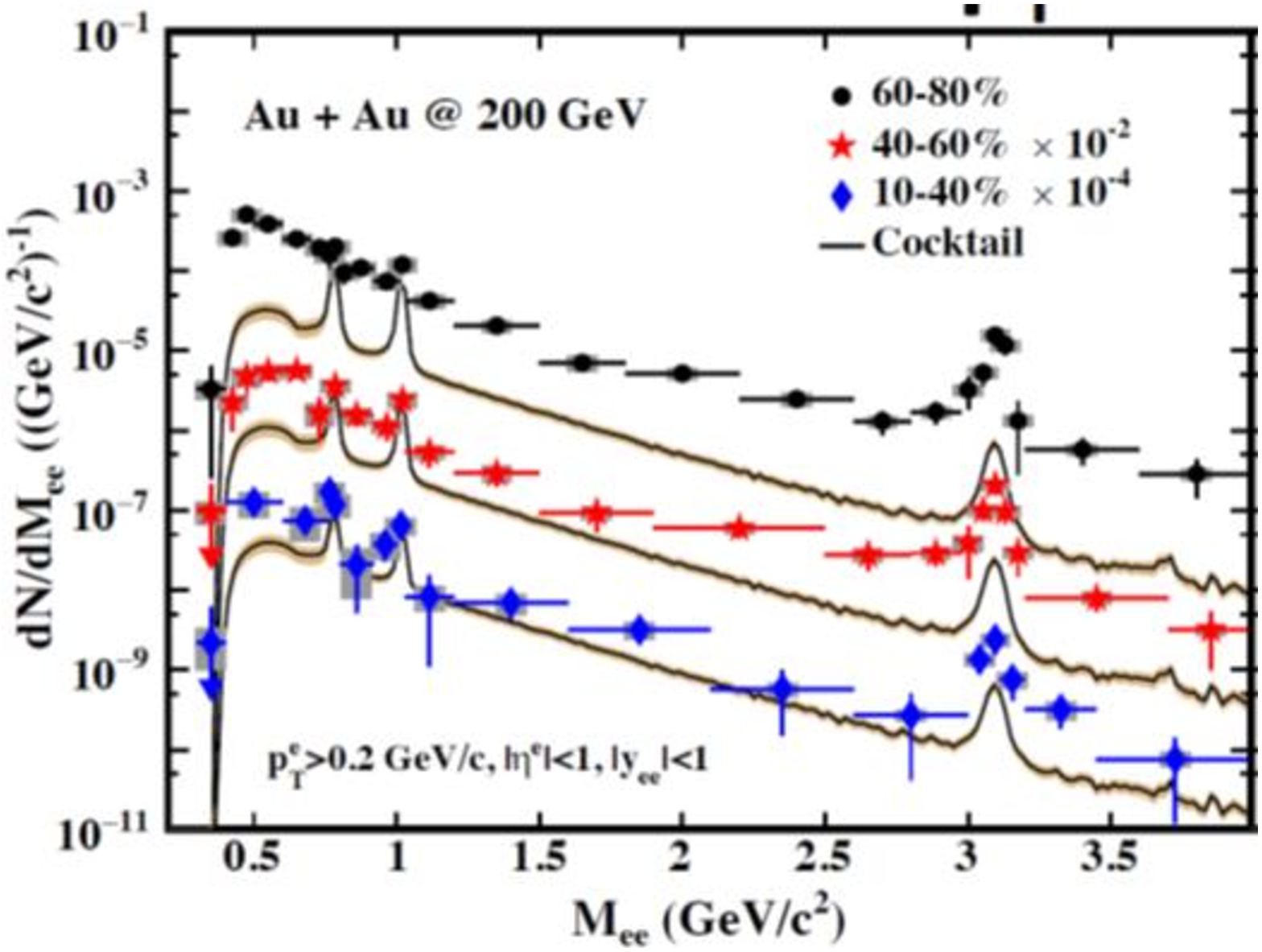}
  \label{Fig:STARee_lowpt}
\end{minipage}
\hfill
\begin{minipage}[t]{7.2cm}
  \centering
  \includegraphics*[width=7.2cm,trim = 1cm 1.2cm 0.5cm 0cm]{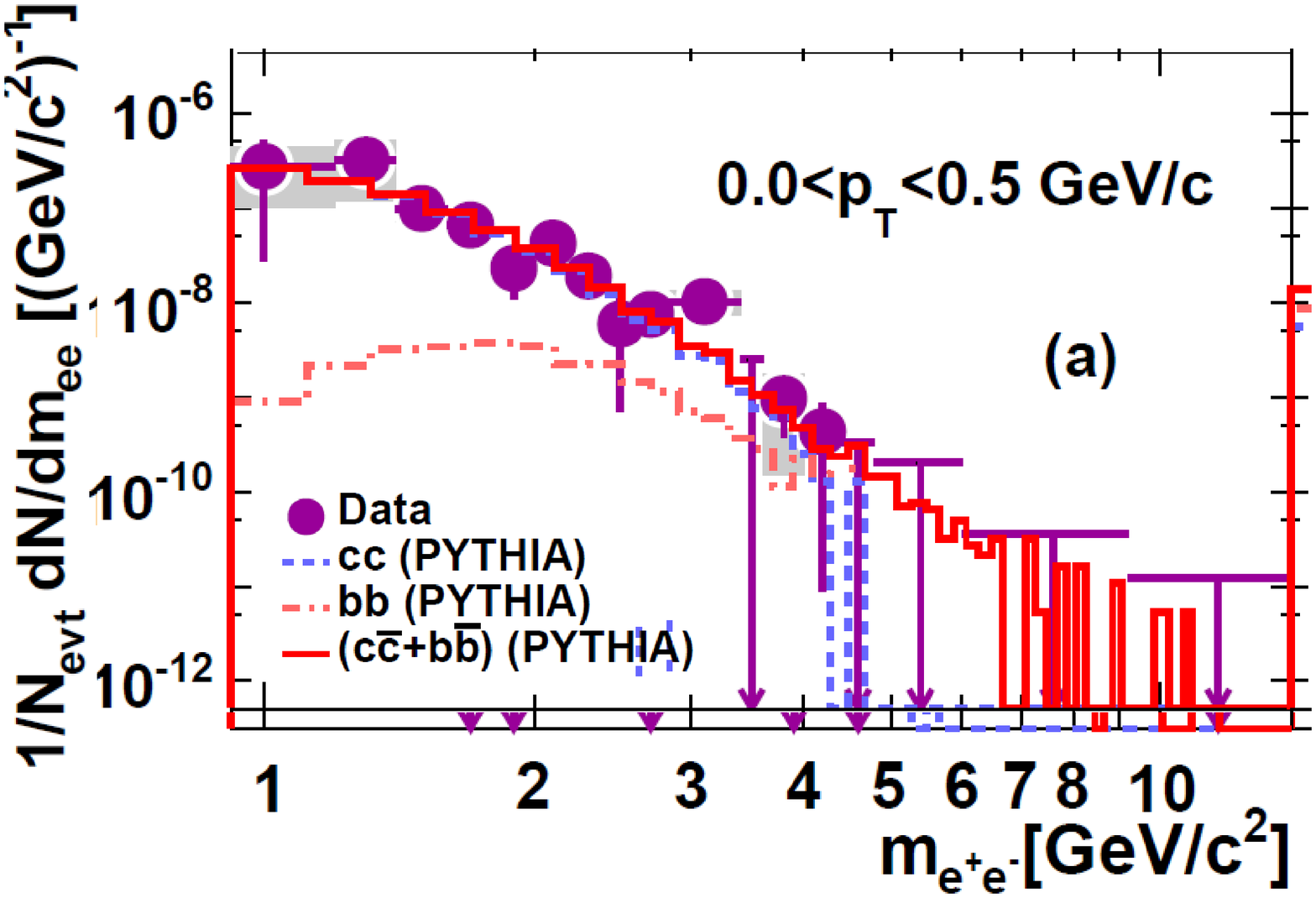}
  \label{Fig:PHENIXee_lowpt}
\end{minipage}
\caption{On the left are preliminary results from the STAR collaboration of the dielectron mass spectra in $\sqrt{s_{NN}}=200$~GeV $Au$+$Au$ collisions.  These spectra are for centrality bins $10-40\%$ (blue), $40-60\%$ (red), and $60-80\%$ (black) in the low $p_{T}$ range, $p_{T} < 0.15$~GeV/$c$.  The lines display the expected cocktail contributions for each centrality bin.  A clear excess at all masses is seen in each centrality bin with the excess more evident in more peripheral bins.  For comparison, on the right, are published results from the PHENIX Collaboration of the dielectron mass spectrum in $\sqrt{s_{NN}}=200$~GeV $d$+$Au$ collisions.~\cite{PHENIXdAudie}  The minimum bias $d$+$Au$ data are shown in the low $p_{T}$ range, $p_{T} < 0.5$~GeV/$c$, and at masses above $1$~GeV/$c^{2}$.  The J/$psi$ and other hadronic contributions have been removed from the $d$+$Au$ spectra.  In the $d$+$Au$ low $p_{T}$ mass spectra no excess is seen and the spectra matches the $PYTHIA$ contributions from correlated open heavy flavor decays.}
\label{Fig:ee_lowpt}
\end{figure}

Newly published heavy flavor cross-section measurements in $\sqrt{s_{NN}}=200$~GeV $p$+$p$ and $d$+$Au$ collisions were shown by the PHENIX experiment.~\cite{PHENIXhf}  The cross-sections are obtained by fitting the dielectron spectrum double differentially in $p_{T}$ and mass.  The fits are performed in the region above the phi meson, $m_{ee} > 1.16$~GeV/$c^2$, where heavy flavor dielectron pairs from semileptonic decays of $c\bar{c}$ and $b\bar{b}$ dominate.  $PYTHIA$, $MC@NLO$ and $POWHEG$ Monte Carlo generators are used to simulate the $c\bar{c}$, $b\bar{b}$ and Drell Yan distributions.  Each of these generators are able to describe the data.  While there is significant systematic variation in the resulting cross-sections obtained from the different Monte Carlo generators, particularly for the $\sigma_{c\bar{c}}$, the $R_{dA}$ for $c\bar{c}$ and $b\bar{b}$ pairs agree well.  The $R_{dA}$ values for dileptons correlated from $c\bar{c}$ and $b\bar{b}$ semileptonic decays are consistent with one showing no evidence for modifications of heavy flavor production in $d$+$Au$ collisions.

Future heavy flavor analyses of the dilepton continuum may benefit from the use of off-vertex pair position as is being investigated by the ALICE experiment.~\cite{ALICEdie}  ALICE showed preliminary results of the dielectron pair distance of closest approach distribution in $\sqrt{s_{NN}}=7$~TeV $p$+$p$ collisions.  The data is compared to Monte Carlo simulations of the expected contribution from $c\bar{c}$ and $b\bar{b}$ semileptonic decays.  It is well described in both the correlated heavy flavor rich region, $1.1 < m_{ee} < 2.7$~GeV/$c^2$, and the region dominated by light flavor meson decays, $0.08 < m_{ee} < 0.14$~GeV/$c^2$.

The ALICE and STAR experiments presented projections of expected future dielectron measurements with improved analysis techniques, such as the use of machine learning algorithms, and upgraded detectors, including STAR's Muon Telescope Dectector upgrade and their longer term iTPC upgrade and the upgrade to the ALICE collaboration's Inner Tracking System and TPC.  Additionally, there is substantial community involvement on future dielectron continuum measurements at low collision energies, $\sqrt{s_{NN}}$ between 3 and 10~GeV, that reach high net baryon densities.   This includes the second beam energy scan at RHIC with the STAR detector, CBM at FAIR and new facilities at NICA and J-PARC.

\section{Direct photon results}
\label{Sec:DirPhot}

The PHENIX~\cite{PHENIXphot1}, STAR~\cite{STARphot} and ALICE~\cite{ALICEphot} experiments have measured excess direct photon production at low $p_{T}$ in heavy ion collisions relative to NLO pQCD calculations or $p$+$p$ results scaled by the number of binary collisions, in $\sqrt{s_{NN}}=200$~GeV $Au$+$Au$ at RHIC and $\sqrt{s_{NN}}=2.76$~TeV $Pb$+$Pb$ collisions at the LHC.  There is a disagreement in the amount of the direct photon yields measured by STAR and PHENIX in the $1$-$3$ GeV/$c$ $p_{T}$ range, with PHENIX reporting larger yields.~\cite{STARphot}  Furthermore, these excess direct photons have substantial azimuthal anisotropies, $v_{n}$, with respect to the collision's event planes.~\cite{PHENIXphotv2}~\cite{ALICEphotv2}  The measured direct photon $v_{n}$ are as large as the $v_{n}$ measured in hadrons.  Theoretical models have difficulty reproducing the large low $p_{T}$ yields with the large $v_{n}$ values.  This is often called the direct photon $v_{n}$ puzzle.  One way to reconcile the direct photon $v_{n}$ puzzle is to have increased photon production late in the collision after the anisotropic pressure gradients have developed, such as coalescence-like production at hadronization.~\cite{Sarah}

An important goal of direct photon measurements is to extract the temperature of the QGP and other collision stages.  By fitting the direct photon excess spectrum with an exponential the inverse slope or effective temperature, $T_{Eff}$, can be determined.  However, the large direct photon anisotropies suggest that the direct photon spectra is boosted or blue-shifted by the strong radial flow of the fireball, resulting in an extracted $T_{Eff}$ that is larger than the true fireball temperature.  In order to determine the true system temperature from the direct photon spectrum, experimentalists must rely on theorists to correctly model the various direct photon contributions.

The direct photon spectra are measured with multiple techniques including calorimeter cluster measurements and the correlations of dielectron pairs produced by the internal conversion of virtual photons or the conversion of real photons in detector material.  A key difficulty of these measurements is to remove the substantial backgrounds from hadronic decay photons.  With dilepton conversion analyses, a double ratio statistical subtraction technique is used to reveal the direct photon contribution to the inclusive photon spectrum.  Equation~\ref{Eq:photdir} shows how the direct photon yield is extracted from the inclusive photon yield by subtracting the decay contributions. Instead of a straight subtraction, the $R_{\gamma}$ double ratio is used to reduce systematic errors because some of the systematic errors cancel in the ratio.  Equation~\ref{Eq:Rgamma} presents how $R_{\gamma}$ is calculated.
\begin{equation}
\gamma_{direct} = \gamma_{incl} - \gamma_{decay} = \left( 1 - \frac{1}{R_{\gamma}}\right) \gamma_{incl}
\label{Eq:photdir}
\end{equation}
\begin{equation}
R_{\gamma} = \frac{\gamma_{incl}}{\gamma_{decay}} = \frac{ \left( \frac{\gamma_{incl}}{\pi^0}\right)_{meas}}{\left(\frac{\gamma_{incl}}{\pi^0}\right)_{sim}}
\label{Eq:Rgamma}
\end{equation}
In real photon conversion analyses, PHENIX uses neutral pion-tagged events to measure the $R_{\gamma}$, where both the $\gamma_{incl}$ and $\pi^0$ yields are measured in events with a tagged $\pi^0$.  While this reduces the statistics, it allows for the cancelation of the photon conversion probability.  ALICE does not perform a neutral-pion tagged analysis, instead relying on their determination of the conversion probability through simulations and including the error on the conversion probability in their systematic error bars.   At the LHC isolated photon spectra are also measured to remove the substantial contributions from fragmentation photons at high collision energies.  ALICE presented a preliminary isolated photon spectrum in $\sqrt{s_{NN}}=7$~TeV $p$+$p$ collisions using their electromagentic calorimeter, EMCal.~\cite{ALICEnewphot}  This measurement extends the isolated photon spectrum down to $10$~GeV in $E_{T}$.  This may allow for an overlap of experimental measurements in different detectors and with different measurement techniques.

\begin{figure}
\centering
\begin{minipage}[t]{7.2cm}
  \centering
  \includegraphics*[width=7cm]{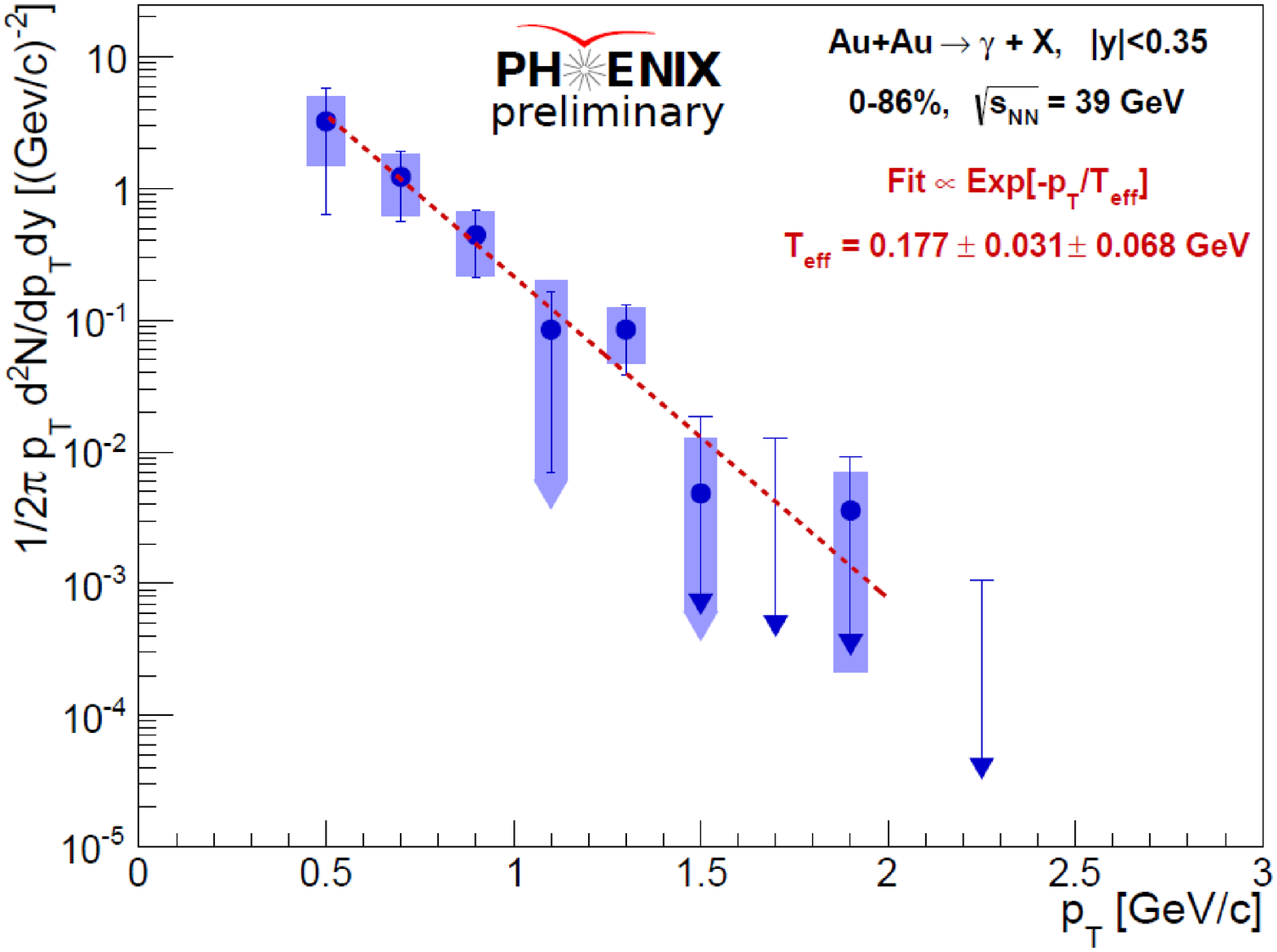}
  \label{Fig:PHENIXdirphot_39}
\end{minipage}
\hfill
\begin{minipage}[t]{7.2cm}
  \centering
  \includegraphics*[width=7.cm]{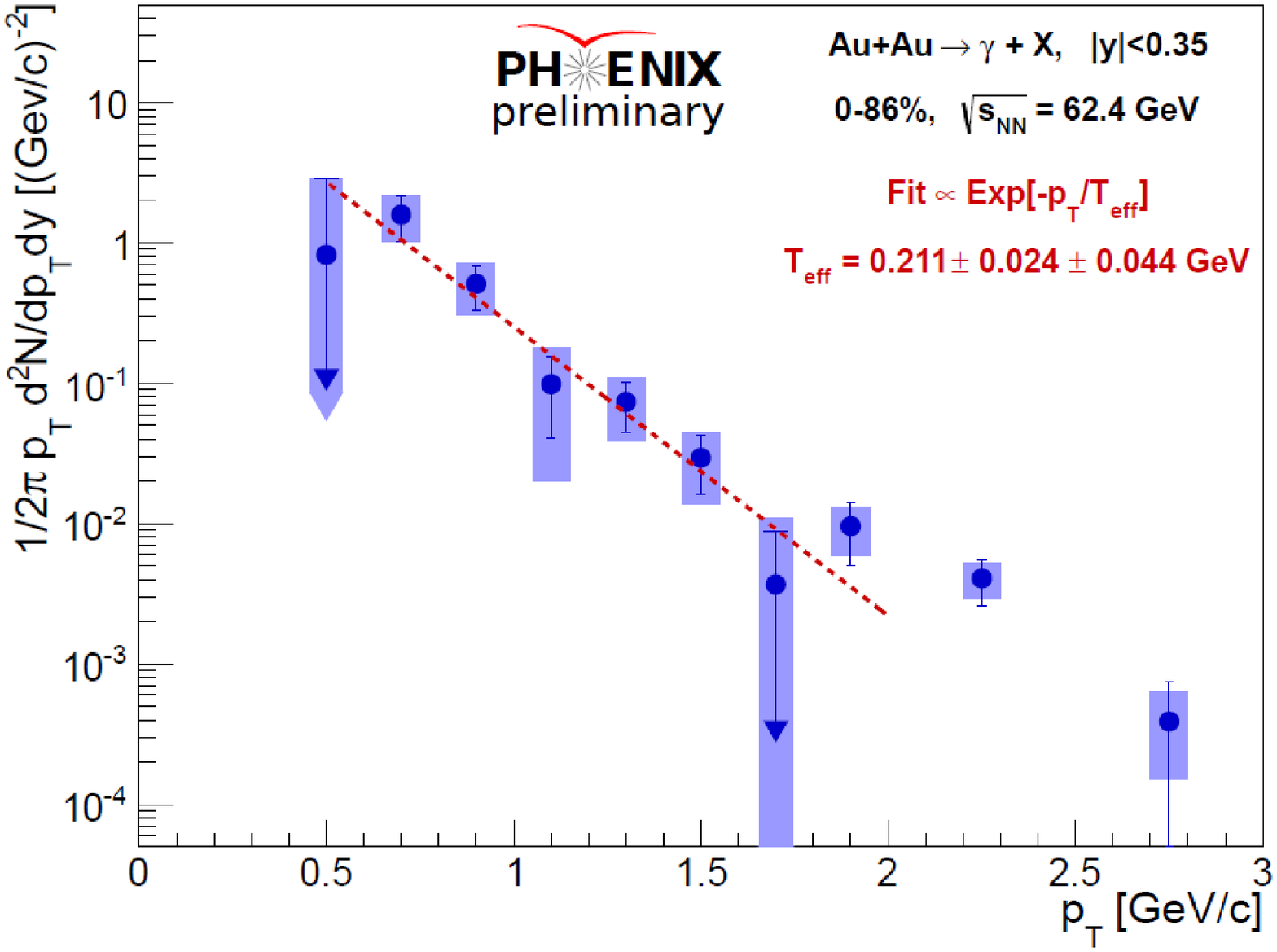} 
  \label{Fig:PHENIXdirphot_62}
\end{minipage}
\caption{PHENIX preliminary measurements of the direct photon spectra in minimum bias $\sqrt{s_{NN}}=39$~GeV (left) and $\sqrt{s_{NN}}=62.4$~GeV (right) $Au$+$Au$ collisions.  These were measured using real photon conversions with the $R_{\gamma}$ statistical subtraction technique.  Exponential fits extracting inverse slope values, $0.177 \pm 0.031 \pm 0.068$~GeV and $0.211\pm0.024\pm0.044$~GeV for $\sqrt{s_{NN}}=39$~GeV and $\sqrt{s_{NN}}=62.4$~GeV respectively, are shown as red dashed lines.}
\label{Fig:PHENIXdirphot_snn}
\end{figure}

PHENIX presented new preliminary results investigating the collision energy and system size dependence of direct photon production.~\cite{Deepali}  They measured the direct photon spectra in $\sqrt{s_{NN}}=39$ and $62.4$~GeV $Au$+$Au$ collisions using real photon conversion analyses.  The minimum bias spectra at both collision energies is shown in Figure~\ref{Fig:PHENIXdirphot_snn}.  The $\sqrt{s_{NN}}=62.4$~GeV $Au$+$Au$ results in $0-20\%$ and $20-40\%$ centrality bins were also shown.  Additionally, preliminary direct photon spectra in $\sqrt{s_{NN}}=200$~GeV $Cu$+$Cu$ collisions using a virtual photon conversion analysis were presented in the minimum bias and the $0-40\%$ centrality bin.

\begin{figure}
    \begin{center}
        \includegraphics[width=9cm]{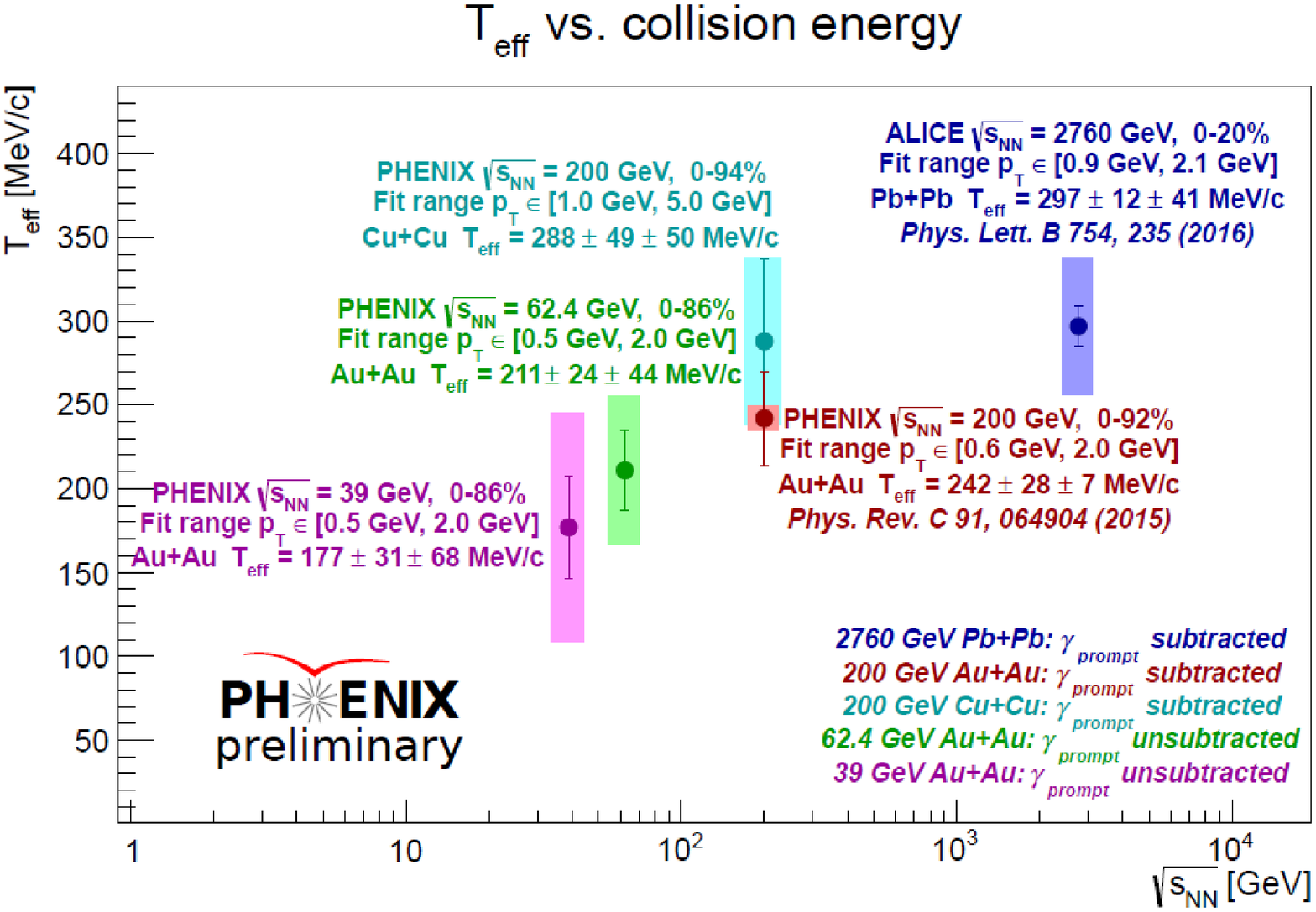}
        \caption{The inverse slope, $T_{Eff}$, as a function of the collisions energy, $\sqrt{s_{NN}}$, for different collision systems.  The inverse slopes are obtained from exponential fits to the direct photon $p_{T}$ spectra at low $p_{T}$ where the excess dominates.  Results obtained by the PHENIX experiment in $Au$+$Au$ collisions at $200$, $62.4$, and $39$~GeV, in $Cu$+$Cu$ collisions at $200$~GeV, and ALICE measurements in $Pb$+$Pb$ collisions at $2.76$~TeV.  Slight increase in $T_{Eff}$ with increasing $\sqrt{s_{NN}}$ is seen.}
        \label{Fig:TeffSnn}
    \end{center}
\end{figure}

With these new results and in conjunction with previously measured $\sqrt{s_{NN}}=200$~GeV $Au$+$Au$ results at PHENIX and $\sqrt{s_{NN}}=7$~TeV $Pb$+$Pb$ results at ALICE, PHENIX compares the direct photon excess yields and the inverse slopes of the $p_{T}$ spectra across collision systems and collision energies. They find that the direct photon yields as a function of the number of nucleons participating the collision, $N_{Part}$, are well described by a fit to $A N_{Part}^\alpha$ with an $\alpha$ factor of $1.36\pm0.09$.  This means that the direct photon yield increases at a rate faster than $N_{Part}$.  However, the similarity of the $62.4$ and $200$~GeV $Au$+$Au$ yields at the same $N_{Part}$ suggests that system size may also play a role.  The inverse slopes, $T_{Eff}$, from exponential fits to the excess direct photon $p_{T}$ spectra are shown as a function of collision energy in Figure~\ref{Fig:TeffSnn}.  A hint of an increase in $T_{Eff}$ with increasing $\sqrt{s_{NN}}$ is seen.  While it is tempting to think of these inverse slope values as related to the various system temperatures, it is important to remember that the anisotropic nature of photon production suggests that the spectra is boosted by the large amount of radial flow as the collision expands. This would result in $T_{Eff}$ values larger than the true fireball temperature.  The increase in $T_{Eff}$ as a function of collision energy, though slight, may in fact be related to the increase in the average boost velocity seen in blast wave analyses of charged particle spectra at various collision energies.

Analysis of the direct photon $v_{2}$ is underway in the PHENIX experiment using both a calorimeter-based method and dielectrons from real photon conversions in detector material.  The direct photon $v_{n}$ is also measured statisictally from the inclusive photon anisotropy according to Equation~\ref{Eq:photv2}, where $R_{\gamma}$ is the same $R_{\gamma}$ defined in Equation~\ref{Eq:Rgamma}.
\begin{equation}
v_{2}^{\gamma, direct} = \frac{R_{\gamma}v_{2}^{\gamma, incl} - v_{2}^{\gamma, decay}}{R_{\gamma}-1}
\label{Eq:photv2}
\end{equation}
Preliminary results of the inclusive photon $v_{2}$ in $0-20\%$ $\sqrt{s_{NN}}=200$~GeV $Au$+$Au$ collisions were shown using the calorimeter and real photon conversion techniques with the 2014 dataset.  The results from both measurement techniques agree.  Future direct photon measurements in $\sqrt{s_{NN}}=200$~GeV $p$+$p$, $p$+$Au$ and $Cu$+$Au$ collisions are expected from PHENIX using the real photon conversion technique.  The ALICE collaboration first presented a preliminary measurement of the direct photon $v_{2}$ in 2012.~\cite{ALICEphotv2}  They expect to produce a publication of this measurement in the future.~\cite{ALICEnewphot}

\section{Conclusion}
\label{Sec:Concl}
These proceedings detail a summary of recent measurements of dileptons and direct photons presented at Quark Matter 2017.  Of note, the excess in the dilepton continuum is still well-described by the broadened rho scenario.  The collision energy dependence of this model will be tested with precise beam energy scan data from the STAR experiment at RHIC.  HADES presented the dielectron spectra in $\sqrt{s_{NN}}=2.4$~GeV $Au$+$Au$ collisions and found that a coarse grained transport model with rho-baryon coupling works well in this baryon-rich regime.  STAR discovered a low $p_{T}$ peripheral excess in $\sqrt{s_{NN}}=200$~GeV $Au$+$Au$ collisions and proposed that these pairs are a result of photo-production.  Comparison with low $p_{T}$ peripheral dilepton measurements in heavy ion collisions from PHENIX and the LHC experiments is needed.  The theory community still struggles to simultaneously describe the excess direct photon $p_{T}$ spectra and azimuthal anisotropies.  Increased late-stage photon production is one way to address this puzzle.  PHENIX finds that the direct photon yield increases faster than the number of nucleons participating in the collision.  Another consequence of the anisotropic direct photon production is that the measured inverse slopes or effective temperatures, $T_{Eff}$, are not true temperatures as the spectra are shifted by radial boost effects.  New PHENIX results show the hint of an increase in $T_{Eff}$ with collision energy.  This is most likely related to an increase in the boost velocity with $s_{NN}$.  Future work for both dilepton continuum physics and direct photons involves extending these measurements to different collision systems and energies, with particular interest in lower collision energies where the net baryon density is higher.

\section{Acknowledgements}
\label{Sec:Ackno}
I thank the conference organizers for coordinating such a interesting conference and providing me with the opportunity to give this talk.  I also would like to thank those who discussed this subject with me as I prepared this talk and specifically the HADES, ALICE, STAR, and PHENIX collaborations for providing us all with these stimulating results.  This work is supported by the U. S. Department of Energy through grant DE-FG-02-86ER40281.






\end{document}